\documentclass[%
twocolumn,
superscriptaddress,
 amsmath,amssymb,
 aps,
prl,
]{revtex4-2}

\usepackage{xcolor} 
\usepackage{graphicx}
\usepackage{dcolumn}
\usepackage{bm}

\usepackage{color}

\usepackage{titlesec}
\usepackage{textcase}
\usepackage{mfirstuc}
\usepackage{amsthm}
\usepackage{mathtools}
\usepackage{multirow}
\usepackage{textcomp}
\usepackage{float}
\usepackage[normalem]{ulem}
\usepackage{dsfont}
\usepackage{mathrsfs}
\usepackage{url}

\begin{document}

\preprint{APS/123-QED}

\title{Asymmetric Floquet-Engineered Mode Coupling in Hybrid Magnonics}

\author{Amin Pishehvar}
\affiliation{ 
    Department of Electrical and Computer Engineering, Northeastern University, Boston, MA 02115, USA
}

\author{Jayakrishnan M. P. Nair}
\affiliation{ 
    Department of Physics, Boston College, 140 Commonwealth Avenue, Chestnut Hill, MA, 02467, USA
}

\author{Zixin Yan}
\affiliation{ 
    Department of Electrical and Computer Engineering, Northeastern University, Boston, MA 02115, USA
}

\author{Yu Jiang}
\affiliation{ 
    Department of Electrical and Computer Engineering, Northeastern University, Boston, MA 02115, USA
}

\author{Benedetta Flebus}
\affiliation{ 
    Department of Physics, Boston College, 140 Commonwealth Avenue, Chestnut Hill, MA, 02467, USA
}

\author{Xufeng Zhang}
\email{xu.zhang@northeastern.edu}
\affiliation{ 
    Department of Electrical and Computer Engineering, Northeastern University, Boston, MA 02115, USA
}
\affiliation{ 
    Department of Physics, Northeastern University, Boston, MA 02115, USA
}

\date{\today}

\begin{abstract}
In hybrid magnonic systems, linear magnon--photon hybridization inherently produces symmetric, reciprocal interactions, precluding asymmetric mode coupling. Floquet driving can tailor mode coupling strengths but, with single-tone modulation, inevitably generates a symmetric interaction that preserves this reciprocity. Here we introduce dual-tone Floquet modulation to unlock a new degree of freedom in hybrid magnonic systems, where the relative phase $\theta$ of two commensurate drives continuously controls the asymmetry of the Floquet-engineered interaction, enabling asymmetric mode coupling absent in existing hybrid magnonic systems. We demonstrate this in a strongly coupled cavity magnonic device, where tuning $\theta$ reversibly switches single-sided Autler--Townes splitting between the two hybrid modes---a direct spectroscopic signature of phase-programmable asymmetric coupling. This approach opens a new path toward controllable nonreciprocal and topological functionalities in hybrid magnonic systems, with broad implications for advanced quantum and classical signal processing.
\end{abstract}


\maketitle



Hybrid magnonics---the coherent coupling of magnons to microwave photons, optical photons, phonons, and superconducting qubits---has emerged as a versatile platform for coherent information processing and quantum science \cite{zarerameshti_magnonics_PhysRep_2022, lachance-quirion_hybridquantum_APEx_2019, li_hybridmagnonics_JAP_2020, awschalom_engineering_IEEETransQE_2021,chumak_spinwave_IEEETransMag_2022,flebus_roadmap_JPCM_2024,zhang_materials_MatTodayElec_2023}. The strong and widely tunable magnon–photon coupling \cite{soykal_strongfield_PRL_2010, huebl_cooperativity_PRL_2013, tabuchi_hybridizing_PRL_2014, zhang_strongcoupling_PRL_2014, goryachev_cavityqed_PRAppl_2014, bai_spinpumping_PRL_2015, tabuchi_coherentcoupling_Science_2015, li_strongcoupling_PRL_2019, hou_nanomagnet_PRL_2019} has enabled landmark demonstrations including qubit–magnon entanglement \cite{lachance-quirion_singlemagnon_Science_2020, lachance-quirion_resolving_SciAdv_2017}, magnon squeezing \cite{weng_magnonsqueezing_NatComm_2026}, non-Hermitian and topological phenomena \cite{harder_dissipative_PRL_2018, zhang_exceptionalpoint_NatComm_2017, zhang_exceptionalsurface_PRL_2019}, and microwave-to-optical transduction \cite{hisatomi_bidirectional_PRB_2016}. These achievements all rest on linear, reciprocal magnon–photon hybridization; yet the most compelling applications---on-chip signal isolation, topological protection, and synthetic gauge fields---demand asymmetric mode coupling that existing hybrid magnonic systems cannot provide.

\begin{figure}[b]
\includegraphics[width=\linewidth]{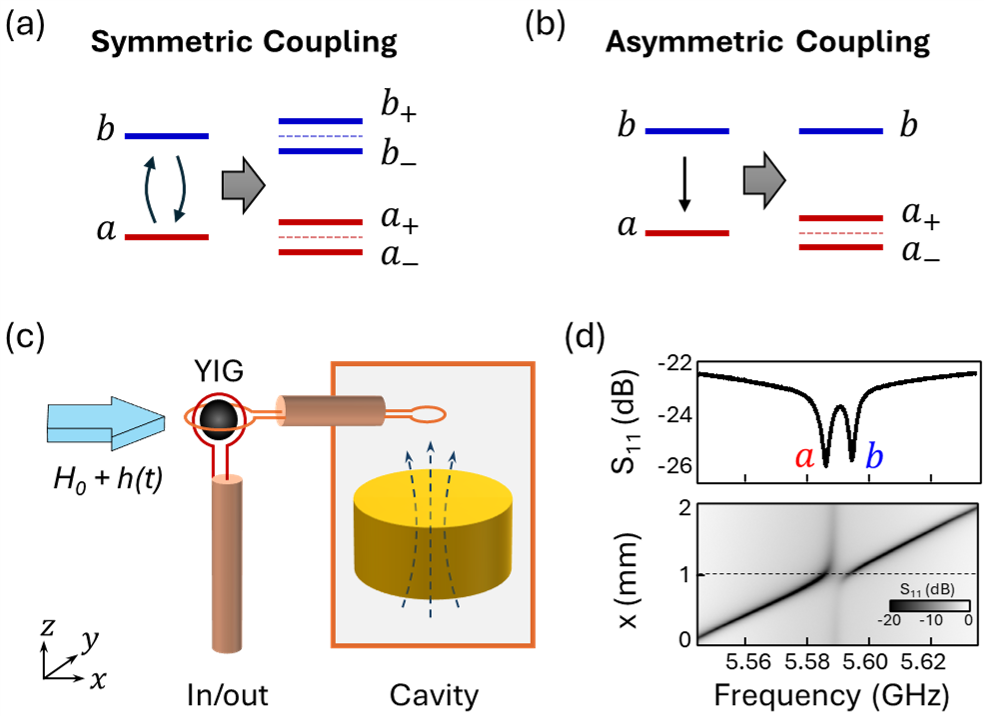}
\caption{(a) Single-tone Floquet driving generates symmetric replicas, leading to equal-strength Floquet-assisted coupling between the hybrid modes $a$ and $b$ and hence ATS on both modes symmetrically. (b) Dual-tone driving with relative phase $\theta$ produces asymmetric Floquet replicas, imbalancing the two opposite Floquet-assisted processes and yielding single-sided ATS predominantly on mode $a$. (c) Schematic of the experimental device. (d) Reflection spectrum $S_{11}$ versus magnet position $x$ (bottom panel), showing resolved normal mode splitting ($g/2\pi \approx 5$~MHz). Top panel shows the spectrum at the dashed line.}
\label{fig1}
\end{figure}

Time-periodic (Floquet) modulation offers a natural route to break this symmetry by tailoring the interaction channels between modes, as demonstrated in photonics \cite{fang_effectivefield_NatPhys_2012, tzuang_nonreciprocal_NatPhoton_2014, dutt_bandstructure_NatComm_2019,dutt_synthdimension_Science_2020}, optomechanics \cite{fang_nonreciprocity_NatPhys_2017}, and superconducting circuits \cite{lee_propagation_PRA_2020} where dynamic modulation has enabled controlled coupling pathways and synthetic gauge effects. 
In hybrid magnonics, modulating the magnon frequency generates Floquet replicas that mediate additional interaction channels between modes \cite{xu_floquet_PRL_2020, pishehvar_floquet_PRAppl_2025, pishehvar_isolation_PRAppl_2025}, establishing Floquet engineering as a viable framework for this platform. Achieving asymmetric coupling within this framework requires the ability to continuously tune the balance of the Floquet-engineered interaction---favoring one Floquet-mediated process over the other---a degree of control that has remained out of reach in existing Floquet magnonic systems, where single-tone modulation produces an inherently symmetric interaction [Fig.~\ref{fig1}(a)].

Here we introduce dual-tone Floquet modulation of the magnon frequency to achieve this control at room temperature. Superposing two commensurate tones produces coherent interference---governed by a tunable relative phase $\theta$---that continuously adjusts the balance of the Floquet-engineered interaction, selectively enhancing one Floquet-mediated process while suppressing the other. Using time-resolved spectroscopy~\cite{dutt_bandstructure_NatComm_2019}, we directly map how $\theta$ reshapes the instantaneous magnon-frequency trajectory, and static reflection spectra confirm the resulting interaction asymmetry. By applying this approach to two hybrid modes in a cavity magnonic device, we directly observe the spectral signature of asymmetric mode coupling, manifested as single-sided Autler--Townes splitting (ATS) [Fig.~\ref{fig1}(b)] that switches continuously with $\theta$.


The two hybrid modes ($a$ and $b$) are realized in the device shown in Fig.~\ref{fig1}(c), which couples a highly polished yttrium iron garnet (YIG) sphere (diameter 400~µm) to a dielectric microwave resonator (diameter 9.5~mm, height 6.0~mm, $\varepsilon_r=30$) housed in a metal enclosure. The YIG sphere is biased by a permanent magnet to support magnon excitation; in this work, the fundamental Kittel mode (denoted by $m$) is used for simplicity. Excitation and readout of the magnon mode are performed via a coaxial loop antenna oriented perpendicular to the bias field. We employ the TE$_{01\delta}$ cavity mode (denoted by $c$) around $\omega_c/2\pi = 5.59$~GHz with a dissipation rate $\kappa_c/2\pi \approx 3$~MHz. Coherent magnon--photon coupling is mediated by a coaxial-link waveguide terminated with two shorted loops: one encircling the YIG sphere and the other located above the dielectric resonator, oriented perpendicular to both the bias field and the readout antenna to maximize coupling while suppressing direct feedthrough.

When the bias field $H_0$ is swept, a clear avoided crossing can be observed in the magnon reflection spectrum [Fig.~\ref{fig1}(d), lower panel], confirming that the system operates in the strong‑coupling regime with $g/2\pi \approx 5$~MHz and magnon linewidth $\kappa_m/2\pi \approx 4$~MHz. The resulting pair of hybridized normal modes $a$ and $b$ [Fig.~\ref{fig1}(d), upper panel] are separated by $\Delta = 2g$, which will be used to investigate the asymmetric coupling induced by Floquet drive. Floquet modulation is applied via two independent coils wound around the YIG sphere, which can provide efficient magnetic field modulation in the frequency range below 20 MHz.


\begin{figure}[tb]
\includegraphics[width=\linewidth]{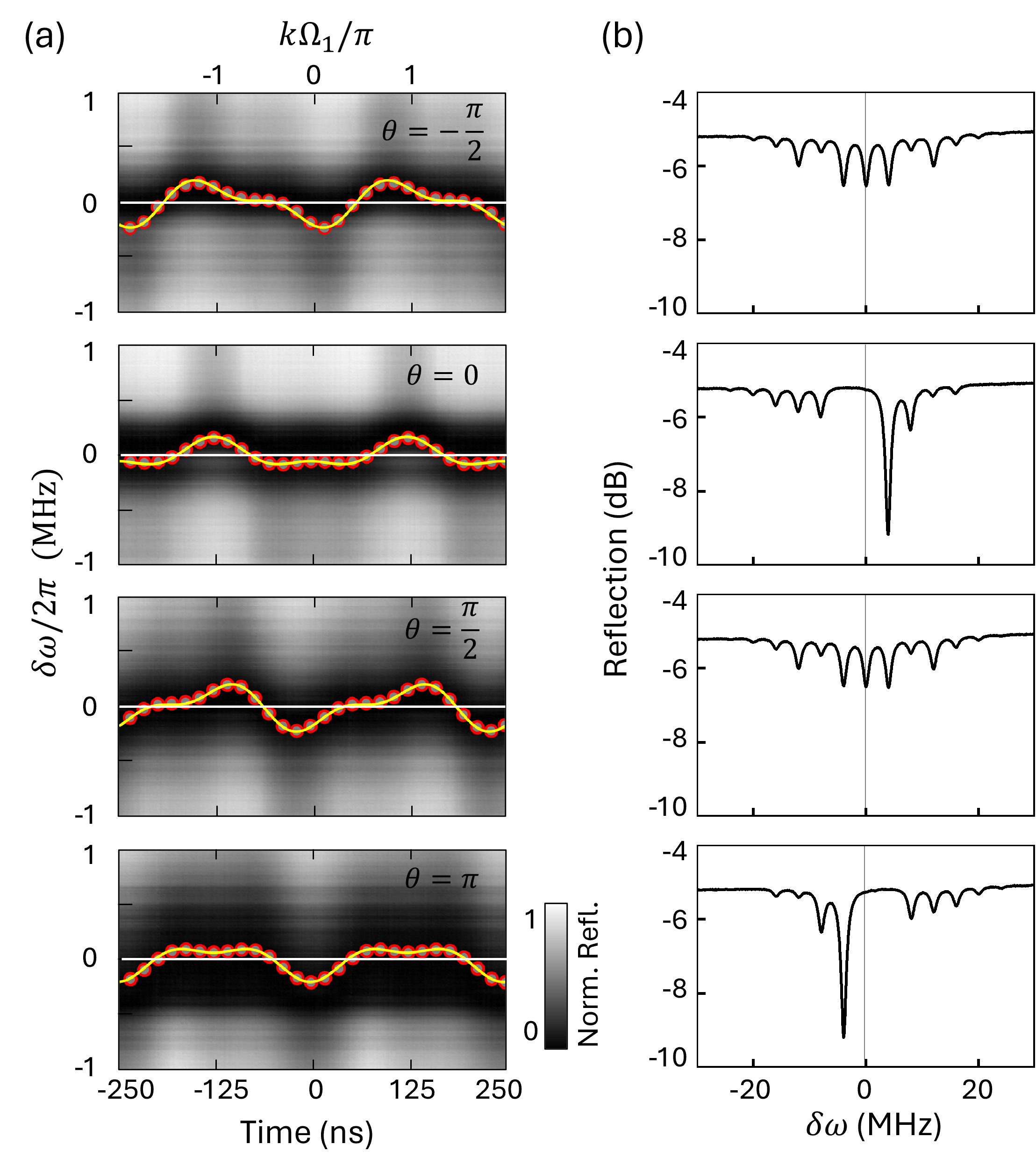}
\caption{(a) Time-resolved reflection spectra of the magnon mode under dual-tone Floquet driving at $\Omega_1/2\pi = 4$~MHz and $\Omega_2/2\pi = 8$~MHz, for drive phases $\theta = -\pi/2$, $0$, $\pi/2$, and $\pi$. Yellow curves show the extracted instantaneous resonance frequency; red curves are numerical calculations based on Eq.~(\ref{Eq:modOmega}). The top axis maps time onto the Floquet quasimomentum $k$. (b) Corresponding static reflection spectra measured at the same drive phases, showing the asymmetric Floquet replica structure in the frequency domain. The vertical line marks $\delta\omega=\omega-\omega_0 = 0$.}
\label{fig2}
\end{figure}

Theoretically, the magnon resonance frequency under a modulation field $h(t)$ becomes $\omega_m(t) = \gamma[H_0 + h(t)]$, where $\gamma$ is the gyromagnetic ratio. A single sinusoidal drive at frequency $\Omega$ always produces symmetric Floquet replicas separated by $\Omega$. To break this symmetry, we superpose two commensurate tones with modulation field $h(t)=h_1\cos(\Omega_1 t)+h_2\cos(\Omega_2 t+\theta)$, yielding
\begin{equation}
    \omega_m(t) = \omega_0 + \delta\omega_1 \cos(\Omega_1 t) + \delta\omega_2 \cos(\Omega_2 t + \theta),
\label{Eq:modOmega}
\end{equation}
where $\omega_0 = \gamma H_0$, $\delta\omega_{1,2} \equiv \gamma h_{1,2}$ are the modulation depths, and $\theta$ is the relative phase between the two drives. Choosing $\Omega_2 = 2\Omega_1$ ensures that the combined modulation has a single fundamental period $T = 2\pi/\Omega_1$, allowing the first- and second-harmonic components to interfere coherently and produce asymmetric coupling between Floquet replicas of the magnon mode.

The dual-tone modulation generates Floquet replicas of the magnon mode labeled by index $n$, corresponding to frequency shifts $n\Omega_1$. The relative amplitudes of these replicas are characterized by complex coefficients $C_n$, given by [see Supplemental Material~\cite{SM}]
\begin{equation}
C_n = \sum_{p=-\infty}^{\infty} J_{n-2p}(\beta_1)\, J_p(\beta_2)\, e^{-i p \theta},
\label{eq2}
\end{equation} 
where $p$ is an integer summation index counting the number 
of $\Omega_2$-drive quanta in each multiphoton pathway 
contributing to the $n$-th Floquet replica, $J_p$ denotes the Bessel function of the first kind, $\beta_1 = \delta\omega_1/\Omega_1$, and $\beta_2 = \delta\omega_2/\Omega_2 = \delta\omega_2/(2\Omega_1)$. Equation \eqref{eq2} shows that the coefficient $C_n$ arises from the coherent superposition of distinct Floquet pathways generated by the two modulation tones: a Floquet replica at frequency shift $n\Omega_1$ can be reached through processes involving $n-2p$ quanta of the $\Omega_1$ drive and $p$ quanta of the $2\Omega_1$ drive, with each pathway weighted by the phase factor $e^{-ip\theta}$. Because these contributions interfere with opposite relative phases for $+n$ and $-n$, dual-tone driving generally breaks the replica symmetry of the Floquet spectrum, yielding unequal spectral weights, $|C_n|^2 \neq |C_{-n}|^2$. 
To leading order in the drive amplitudes, the asymmetry takes the form $(|C_{+n}|^2 - |C_{-n}|^2) \propto \cos\theta$ for the low-order replicas $n = \pm 1$ and $n = \pm 2$ relevant to our experiment, with explicit expressions given in the Supplemental Material~\cite{SM}.
The relative phase $\theta$ therefore controls the contrast between the two opposite replicas continuously, with maximal asymmetry at $\theta = 0, \pm\pi$ and restored symmetry at $\theta = \pm\pi/2$.

In our experiment, this dual-tone modulation is implemented by applying tones at $\Omega_1$ and $\Omega_2=2\Omega_1$ through two independent coils aligned parallel to the static field, with drive amplitudes $V_{1,2}$ that determine the modulation depths as $\delta\omega_{1,2} \propto V_{1,2}$. We first characterize its effect on the magnon mode alone, after removing the cavity and coaxial-link waveguide to isolate the magnonic response. Figure~\ref{fig2}(a) shows time-resolved magnon reflection spectra for four representative drive phases $\theta = -\pi/2, 0, \pi/2$, and $\pi$, plotted as a function of the detuning $\delta\omega=\omega-\omega_0$ and time. The yellow trace marks the extracted instantaneous resonance frequency $\omega_m(t)$; red curves are numerical calculations based on Eq.~(\ref{Eq:modOmega}), which agree closely with the measurements.
 
The trajectory of the magnon resonance frequency $\omega_m(t)$ is directly controlled by $\theta$. At $\theta = \pm\pi/2$, the positive and negative frequency excursions are equal in amplitude; at all other phases the trajectory becomes asymmetric about $\delta\omega=0$, with the positive and negative excursions differing in amplitude. Such asymmetry reaches maximum at $\theta = 0$---where the excursion toward positive detuning dominates---and at $\theta = \pi$, where the asymmetry reverses. Additional phases are shown in the Supplemental Material~\cite{SM}. These measurements demonstrate that $\theta$ provides continuous, full-cycle control over the asymmetry of the magnon frequency trajectory. Such asymmetry corresponds directly to asymmetric Floquet replicas in the frequency domain, which is confirmed by the static reflection spectra measured at the corresponding phases [Fig.~\ref{fig2}(b)].

\begin{figure}[tb]
\includegraphics[width=\linewidth]{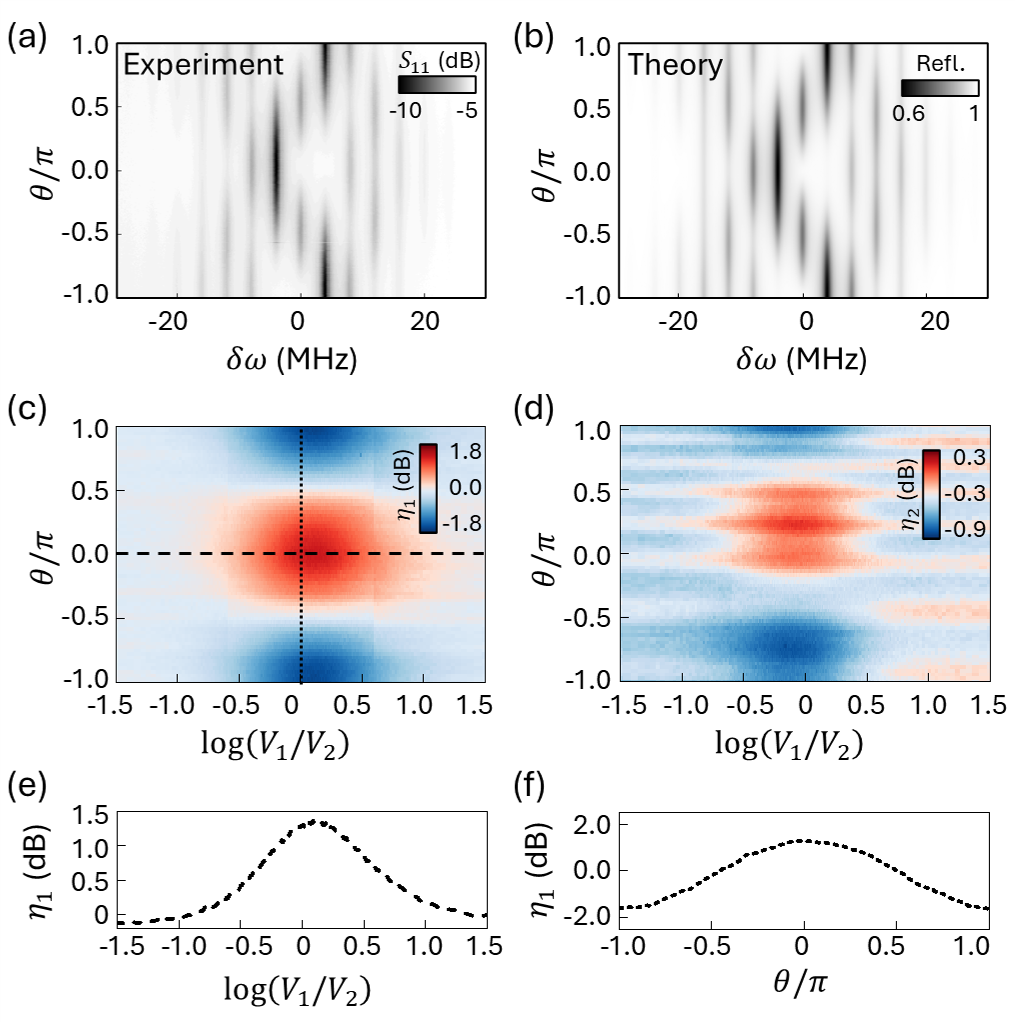}
\caption{(a) Measured and (b) calculated static reflection spectra as a function of drive phase $\theta$, plotted versus detuning $\delta\omega = \omega - \omega_0$. Experiment in (a) uses $V_1 = V_2 = 20$~V; calculation in (b) uses $\delta\omega_1/2\pi = \delta\omega_2/2\pi = 10$~MHz. (c),(d) Measured asymmetry factors $\eta_1$ and $\eta_2$ for the first and second Floquet replicas, plotted as functions of $\theta$ and the voltage ratio $V_1/V_2$. (e) Linecut of (c) along the dashed horizontal line ($\theta = 0$), showing the peaked dependence of $\eta_1$ on $V_1/V_2$ with maximum asymmetry near $V_1 = V_2$. 
(f) Linecut of (c) along the dashed vertical line ($V_1 = V_2$), showing the sinusoidal dependence of $\eta_1$ on $\theta$.} 
\label{fig3}
\end{figure}

Figure~\ref{fig3}(a) examines this phase dependence more systematically, showing the magnon reflection spectra as $\theta$ is swept continuously with equal drive amplitudes $V_1 = V_2 = 20$~V; the overall evolution agrees closely with theoretical calculations [Fig.~\ref{fig3}(b)]. The phase dependence is continuous and controllable: at $\theta = \pm\pi/2$, multiple Floquet replicas appear symmetrically around the magnon resonance, while at $\theta = 0$ the replica structure becomes markedly asymmetric: the lower-frequency replica at $\delta\omega/2\pi = -4$~MHz is strongly enhanced while the upper replica at $\delta\omega/2\pi = 4$~MHz and the central carrier at $\delta\omega = 0$ are extinguished. This behavior reverses at $\theta = \pi$, where the upper replica is enhanced and the lower replica is suppressed.

To quantify this phase-controlled replica asymmetry, we define an asymmetry factor $\eta_n = S_n^+ - S_n^-$, where $S_n^+$ and $S_n^-$ denote the measured reflection amplitudes (in dB) of the upper and lower Floquet replicas of order $n$. The asymmetry depends not only on $\theta$ but also on the relative drive amplitudes $V_1$ and $V_2$. Figures~\ref{fig3}(c),(d) show the measured asymmetry factors $\eta_1$ and $\eta_2$ for the first and second Floquet replicas, plotted as functions of $\theta$ and the voltage ratio $V_1/V_2$. Both replicas exhibit strong and tunable asymmetry, with $\eta_1$ showing the largest contrast. Representative linecuts [Figs.~\ref{fig3}(e),(f)] confirm that the asymmetry varies sinusoidally with $\theta$ and displays a peaked dependence on the drive-amplitude ratio (see Eq.~S17 in Supplemental Material \cite{SM}), reaching maximum asymmetry near $V_1 = V_2$, where the interference contrast between the two modulation harmonics is maximized. These measurements demonstrate that by adjusting only the phase and amplitude of the two modulation tones, one can continuously and selectively enhance or suppress the amplitude of individual Floquet replicas. This, in turn, governs the asymmetry between forward and backward Floquet-mediated coupling, enabling the phase-programmable asymmetric intermode coupling demonstrated below.


\begin{figure}[tb]
\includegraphics[width=\linewidth]{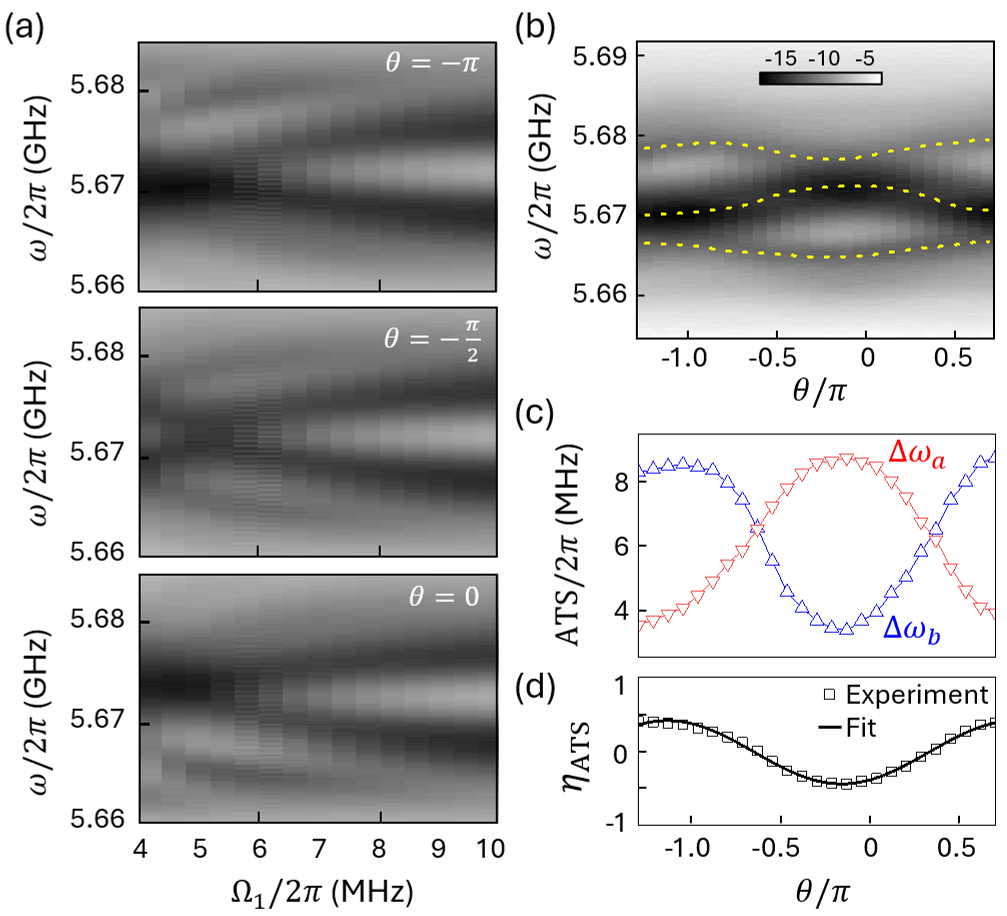}
\caption{(a) Reflection spectra of the hybrid device under dual-tone Floquet driving at $\theta = -\pi$, $-\pi/2$, and $0$, respectively. (b) Reflection spectra as a function of drive phase $\theta$ at $\Omega_1/2\pi = 5$ MHz. Dotted yellow lines indicate the center of the three branches, corresponding to the Floquet hybrid modes (bottom: $a_-$; top: $b_+$; middle: $a_+$ and $b_-$, approximated as identical since their spacing is smaller than the linewidth and thus indistinguishable). (c) Extracted ATS splittings $\Delta\omega_a$ (downward triangles) and $\Delta\omega_b$ (upward triangles) as a function of $\theta$. (d) Extracted $\eta_\mathrm{ATS}$ (squares) and cosine fit (solid line) as a function of $\theta$.}
\label{fig4}
\end{figure}

We now apply this engineered Floquet-path imbalance to demonstrate asymmetric coupling between the two hybrid modes $a$ and $b$ [Fig.~\ref{fig1}(d)], restoring the cavity and coaxial-link waveguide that were removed for the characterization measurements of Figs.~\ref{fig2} and \ref{fig3}. The replica imbalance directly translates into unequal effective coupling strengths for the two intermode transitions (see Supplemental Material~\cite{SM}): in contrast to single-tone driving, which produces symmetric ATS on both modes~\cite{xu_floquet_PRL_2020}, dual-tone driving allows one hybridization channel to be selectively strengthened while the other is suppressed, enabling on-demand control over the symmetry of the coupling.

Figure~\ref{fig4}(a) show the measured reflection spectra at three representative phases. The two hybrid modes are separated by $\Delta/2\pi = 10$~MHz and well resolved in the reflection spectra, allowing clear observation of the ATS---including both symmetric and asymmetric cases---on both modes. In contrast to the magnon-only measurements of Figs.~2 and 3, where the asymmetry is largest at the first-order replicas n=±1, here the coupling between the two hybrid modes is mediated by the $n = 2$ Floquet replica at $\delta\omega/2\pi = 10$~MHz, generated by drives at $\Omega_1/2\pi = 5$~MHz and $\Omega_2 = 2\Omega_1$. 
This second-harmonic replica is used deliberately: the first-harmonic replica would require $\Omega_1/2\pi = 10$~MHz and consequently $\Omega_2/2\pi = 20$~MHz, where the low-pass response of the coils severely suppresses drive amplitude and prevents access to the required $V_1/V_2$ ratio for asymmetric-replica generation.

The results at three selected phases in Fig.~\ref{fig4} (a) reveal distinct coupling behaviors. At $\theta = -\pi/2$, ATS appears symmetrically on both modes, the same as the single-drive case~\cite{xu_floquet_PRL_2020}, indicating balanced coupling between opposite Floquet replicas. At $\theta = 0$, ATS emerges only on the lower mode ($a$), while at $\theta = -\pi$ it appears only on upper mode ($b$). In both cases the coupling becomes effectively asymmetric, indicating that the interaction between the two hybrid modes is predominantly mediated by a single Floquet replica. This behavior originates directly from the phase-dependent asymmetry of the Floquet replica amplitudes, which translates into unequal effective coupling strengths for the two intermode transitions. The corresponding effective replica-assisted cavity--magnon coupling takes the form $\tilde g_n \sim gC_n$ (see Supplemental Material \cite{SM}). Since the two-tone drive generates an asymmetry between $C_n$ and $C_{-n}$, the corresponding couplings mediated by the $+n$ and $-n$ Floquet replicas, $\tilde g_n$ and $\tilde g_{-n}$, become unequal. This imbalance directly produces asymmetric hybridization between the cavity--magnon modes and therefore leads to the observed asymmetry in the ATS. For the experimentally relevant case $n=\pm2$, the condition $|\tilde g_{2}| = |\tilde g_{-2}|$ at $\theta=-\pi/2$ leads to symmetric hybridization of the two modes, while at $\theta=0$ and $\theta=-\pi$, one of the replica-assisted coupling channels becomes strongly suppressed relative to the other, resulting in effectively asymmetric intermode coupling mediated predominantly by a single Floquet replica.

The continuous tunability of the coupling symmetry is further demonstrated by the phase-sweep measurements in Fig.~\ref{fig4}(b). At $\Omega_1/2\pi = 5$~MHz, the ATS alternates periodically from mode $a$ to both modes and then to mode $b$ as $\theta$ is varied. 
To quantify the coupling asymmetry, we extract the ATS splittings $\Delta\omega_a$ and $\Delta\omega_b$ of the two hybrid modes from the phase-sweep data and compute the asymmetry metric defined as $\eta_{\mathrm{ATS}} = (\Delta\omega_a - \Delta\omega_b)/(\Delta\omega_a + \Delta\omega_b)$, shown in Figs.~\ref{fig4}(c) and~\ref{fig4}(d), respectively. The individual splittings oscillate out of phase as $\theta$ is varied [Fig.~\ref{fig4}(c)], directly reflecting the redistribution of spectral weight between the two Floquet-mediated coupling channels. The asymmetry $\eta_{\mathrm{ATS}}$ follows a $\cos\theta$ dependence [Fig.~\ref{fig4}(d)], consistent with the leading-order behavior of $|C_{+2}|^2 - |C_{-2}|^2$, with extrema near $\theta = 0$ and $\theta = -\pi$ (with slight systematic offset) where the coupling is predominantly single-sided, and vanishing near $\theta = -\pi/2$ where both modes exhibit comparable splitting. This provides a quantitative metric of the phase-controlled asymmetric coupling between the two hybrid modes, in direct analogy with the asymmetry factors $\eta_n$ for the magnon-only measurements in Fig.~\ref{fig3}.

In summary, we show that dual-tone Floquet modulation provides direct control over the asymmetry of the Floquet replicas of a magnon mode through the relative amplitude and phase of two commensurate drives. This engineered asymmetry arises from interference between distinct multi-step Floquet pathways, which redistributes spectral weight unevenly between the $+n$ and $-n$ Floquet replicas. As a result, the Floquet-induced interaction between the two magnon--photon hybrid modes becomes phase-tunable: by tuning $\theta$, one coupling channel can be enhanced while the opposite one is suppressed, giving rise to tunable single-sided ATS that switches controllably between the two modes as $\theta$ is varied. These results establish a dynamically programmable route to asymmetric mode interactions in hybrid magnonics, and provide a building block for achieving nonreciprocal wave transport when extended to multi-mode networks. The dual-tone Floquet approach demonstrated here generalizes naturally to other cavity systems supporting bichromatic parametric modulation. In particular, its realization in a hybrid system where two distinct degrees of freedom are strongly coupled opens immediate perspectives for asymmetric Floquet engineering in cavity optomechanics (phonon--photon), circuit QED (qubit--photon), and optomagnonics (magnon--optical photon), at frequencies ranging from microwave to optical.

Future devices incorporating broadband modulation coils or impedance-matched driving networks could access stronger modulation depths and higher-order Floquet replicas. When extended to a multimode magnonic system with equally spaced modes, the Floquet replicas couple adjacent magnon modes in a frequency ladder that constitutes a synthetic dimension. Importantly, the time-resolved data in Fig.~\ref{fig2}(a) also exhibit asymmetry along the time axis; since time maps onto a quasimomentum $k$, such asymmetry---also controlled by $\theta$---implies a broken $k \to -k$ symmetry of the Floquet band structure, which corresponds to a tunable Peierls phase acting as a synthetic gauge flux threading the frequency lattice. This opens a direct route to chiral state transfer and Floquet-engineered topological phases in magnonic arrays. While the present measurements are performed at room temperature, the underlying principle of phase-controlled asymmetric coupling applies equally to the quantum regime, with broad implications for coherent classical and quantum magnonic technologies.


\section{Acknowledgments}
\begin{acknowledgments}
X. Zhang and B. Flebus acknowledge support from National Science Foundation under Grant No. NSF ECCS-2337713.
\end{acknowledgments}

\bibliography{maintext_arxiv}

%

\clearpage
\onecolumngrid

\setcounter{figure}{0}
\setcounter{equation}{0}
\setcounter{section}{0}
\renewcommand{\thefigure}{S\arabic{figure}}
\renewcommand{\theequation}{S\arabic{equation}}
\renewcommand{\thesection}{\arabic{section}}
\renewcommand{\thesubsection}{\thesection.\arabic{subsection}}
\renewcommand{\thesubsubsection}{\thesubsection.\arabic{subsubsection}}
\setcounter{secnumdepth}{3}

\makeatletter
\providecommand{\subsubsection}{}
\renewcommand{\subsubsection}{%
  \@startsection{subsubsection}{3}{0pt}%
    {1.0ex plus .2ex minus .2ex}%
    {0.5ex plus .2ex}%
    {\normalfont\normalsize\bfseries\filcenter}%
}
\makeatother
\titleformat{\section}
  {\normalfont\large\bfseries\filcenter}
  {\thesection}{1em}{\MakeTextUppercase}
\titleformat{\subsection}
  {\normalfont\normalsize\bfseries\filcenter}
  {\thesubsection}{1em}{\capitalisewords}
\titleformat{\subsubsection}
  {\normalfont\normalsize\bfseries\filcenter}
  {\thesubsubsection}{1em}{\capitalizethefirst}

\begin{center}
\textbf{\large Supplemental Material for\\[3pt]
``Asymmetric Floquet-Engineered Mode Coupling in Hybrid Magnonics''}
\end{center}



\section{Theoretical Model}
\subsection{Floquet Green--function formulation of the driven magnon mode}\label{sec.1.1}

In this section we derive the reflection coefficient of a parametrically
modulated magnon mode using a Floquet Green--function approach. We consider a single magnon mode whose resonance frequency $\omega_0$ is modulated by
two harmonic drives at frequencies $\Omega_1$ and $2\Omega_1$.
Including damping and an input field, the Heisenberg--Langevin equation
obtained from the Hamiltonian reads
\begin{equation}
\dot{m}
= -i\!\left[
\omega_0 - i\kappa_m
+ \delta\omega_1 \cos(\Omega_1 t)
+ \delta\omega_2 \cos(2\Omega_1 t + \theta)
\right] m
+ \sqrt{\kappa_{ext}}\, m^{\mathrm{in}} .
\label{eq:eom_time}
\end{equation}
Here, $m$ denotes the magnon annihilation operator, $\omega_0$ represents the bare magnon resonance frequency, while the parameter $\kappa_m$ accounts for the intrinsic damping rate of the magnon mode. 
The parameters $\delta\omega_1$ and $\delta\omega_2$ are the amplitudes of the frequency modulation at the fundamental and second harmonic of the drive, $\Omega_1$ and $2\Omega_1$, respectively, while $\theta$ is their relative phase. The coupling to the external port is quantified by $\kappa_{\mathrm{ext}}$, and $m^{\mathrm{in}}$ denotes the incident probe field. Since the system is driven by a classical input field and remains linear, we work at the level of expectation values and replace the operator $m$ by its coherent amplitude $m(t)=\langle m \rangle$. In this semiclassical limit, quantum noise does not affect the mean dynamics, and Eq.~\eqref{eq:eom_time} reduces to a classical equation for the magnon amplitude.
 Because the Hamiltonian is periodic in time, with period $T = 2\pi/\Omega_1$, we expand the field amplitude in Floquet harmonics as 
\begin{equation}
m(t)
= \sum_{n\in\mathbb{Z}} m_n(t)\, e^{-i n \Omega_1 t} ,
\label{eq:floquet_expansion}
\end{equation}
where the amplitudes $m_n$ correspond to Floquet replicas shifted by integer multiples of the drive frequency ($n\Omega_1$) from the bare magnon resonance frequency $\omega_0$.
 Substituting Eq.~(\ref{eq:floquet_expansion}) into
Eq.~\eqref{eq:eom_time} and collecting equal harmonics $e^{-in\Omega_1 t}$
yields the coupled Floquet equations
\begin{align}
\dot{m}_n
&= -i\!\left[(\omega_0 - n\Omega_1) - i\kappa_m\right] m_n
- i \frac{\delta\omega_1}{2} \left( m_{n-1} + m_{n+1} \right)
\nonumber\\
&\quad
- i \frac{\delta\omega_2}{2}
\left(
m_{n-2} e^{-i\theta}
+ m_{n+2} e^{i\theta}
\right)
+ \sqrt{\kappa_{ext}}\, m^{\mathrm{in}}\, \delta_{n,0} .
\label{eq:floquet_eom}
\end{align}
Equation~\eqref{eq:floquet_eom} maps the dynamics onto a synthetic lattice in frequency space, where each Floquet component \(m_n\) defines a site at energy \(\omega_0 - n\Omega_1\). The first- and second-harmonic modulations generate nearest- and next-nearest-neighbor couplings along this lattice, with the phase \(\theta\) acting as a Peierls phase for the latter. 
To make this structure explicit, we assemble the amplitudes of the Floquet replicas into the Floquet vector
\begin{equation}
\bold{M}
=
\begin{bmatrix}
\cdots & m_{-1} & m_0 & m_1 & \cdots
\end{bmatrix}^{\!T},
\label{eq:floquet_vector}
\end{equation}
so that Eq.~\eqref{eq:floquet_eom} can be written compactly as
\begin{equation}
\dot{\bold{M}}
= -i \bigl[ \mathbf{H} - i\kappa_m\mathbf{I} \bigr] \bold{M}
+ \sqrt{\kappa_{ext}}\, m^{\mathrm{in}}\, \bold{e}_0 ,
\label{eq:matrix_eom}
\end{equation}
where $\boldsymbol{e}_0$ denotes the basis vector in Floquet space with components $(\boldsymbol{e}_0)_n = \delta_{n,0}$. The Floquet dynamical matrix $\mathbf{H}$ has matrix elements given by
\begin{equation}
H_{n n} = \omega_0 - n\Omega_1, \qquad
H_{n,n\pm1}=\frac{\delta\omega_1}{2},
\qquad
H_{n,n+2}=\frac{\delta\omega_2}{2}e^{+i\theta},
\qquad
H_{n,n-2}=\frac{\delta\omega_2}{2}e^{-i\theta}.
\end{equation}
In the frequency domain, the steady-state Floquet amplitudes satisfy
\begin{equation}
\bold{M}(\omega)
= -i
\bigl[
\mathbf{H} - \omega \mathbf{I} - i\kappa_m\mathbf{I}
\bigr]^{-1}
\sqrt{\kappa_{ext}}\, m^{\mathrm{in}}\, \bold{e}_0 .
\label{eq:matrix_solution}
\end{equation}
We define the Floquet Green function
\begin{equation}
\mathbf{G}(\omega)
=
\bigl[
\mathbf{H} - \omega \mathbf{I} - i\kappa_m \mathbf{I}
\bigr ]^{-1}.
\end{equation}
Using the input--output relation
$m^{\mathrm{out}}+m^{\mathrm{in}} = \sqrt{\kappa_{ext}}\,m$,
the reflection coefficient for the probe frequency can be written as
\begin{equation}
r(\omega)
= \sqrt{\kappa_{ext}}\, m_0(\omega)/m^{in} - 1 .
\label{eq:reflection}
\end{equation}
Equation~\eqref{eq:reflection} provides the formal expression for the observed reflection amplitude $|r(\omega)|^2$, with \(m_{0}(\omega)\) obtained using the Floquet Green's function by numerical inversion of the truncated Floquet matrix. 

The resulting spectra, shown in Fig.~3(a) of main text (we truncate Floquet matrix with replica index $N=10$), exhibit a clear redistribution of spectral weight from the central ($n=0$) component into Floquet replicas at $n=\pm 1,\pm 2,\ldots$. Importantly, this redistribution is not symmetric: the amplitudes of the positive- and negative-order replicas differ, and their relative weights depend sensitively on the phase $\theta$ of the two-tone modulation. To clarify the origin of this imbalance, we temporarily neglect the damping $\kappa_m$ and the input field $m^{\mathrm{in}}$, and isolate the effect of the periodic phase modulation induced by the time-dependent magnon frequency. In this limit, the formal solution to Eq.~(\ref{eq:eom_time}) can be written as 
\begin{equation}
m(t)= m(0)\exp\!\left[
-i\omega_0 t
-i\beta_1\sin(\Omega_1 t)
-i\beta_2\sin(2\Omega_1 t+\theta)
\right],
\end{equation}
where $\beta_1 = \delta\omega_1 / \Omega_1$ and $\beta_2 = \delta\omega_2 / (2\Omega_1)$. Expanding the phase factor in Floquet harmonics,
\begin{equation}
e^{-i \epsilon(t)}
=
\sum_{n} C_n \, e^{-i n \Omega_1 t}, \label{s14}
\end{equation}
with $\epsilon(t)=\beta_1 \sin(\Omega_1 t)+\beta_2 \sin(2\Omega_1 t + \theta)$ yields
\begin{equation}
C_n
=
\sum_{p}
J_{n-2p}(\beta_1)\,
J_p(\beta_2)\,
e^{-i p \theta},
\end{equation}
where $J_n$ denotes the Bessel function of the first kind. 
The coefficient $C_n$ characterizes the spectral weight of the $n$-th Floquet replica. In general, $|C_{+n}| \neq |C_{-n}|$, and this imbalance depends explicitly on the phase $\theta$. Focusing on the illustrative case $n=\pm 2$, the leading contributions are
\begin{align}
C_{+2}
&\simeq
J_2(\beta_1)\, J_0(\beta_2)
+
J_0(\beta_1)\, J_1(\beta_2)\, e^{-i\theta}, \\
C_{-2}
&\simeq
J_2(\beta_1)\, J_0(\beta_2)
-
J_0(\beta_1)\, J_1(\beta_2)\, e^{i\theta}.
\end{align}
Both replicas receive contributions from the same two elementary processes: a two-step process mediated by the $\Omega_1$ modulation, and a direct process arising from the $2\Omega_1$ modulation. However, these contributions interfere with opposite relative signs in the $+2$ and $-2$ channels, leading to unequal spectral weights in the positive and negative Floquet sectors:
\begin{align}
|C_{+2}|^2
&\approx
J_2^2(\beta_1)\,J_0^2(\beta_2)
+
J_0^2(\beta_1)\,J_1^2(\beta_2)
+
2\,J_2(\beta_1)\,J_0(\beta_2)\,J_0(\beta_1)\,J_1(\beta_2)\cos\theta,
\\
|C_{-2}|^2
&\approx
J_2^2(\beta_1)\,J_0^2(\beta_2)
+
J_0^2(\beta_1)\,J_1^2(\beta_2)
-
2\,J_2(\beta_1)\,J_0(\beta_2)\,J_0(\beta_1)\,J_1(\beta_2)\cos\theta.
\end{align}
Hence
\begin{equation}
|C_{+2}|^2 - |C_{-2}|^2
=
4\,J_2(\beta_1)\,J_0(\beta_2)\,J_0(\beta_1)\,J_1(\beta_2)\cos\theta ,\label{s20}
\end{equation}
which shows that the imbalance is set entirely by the interference term: it is maximal at \(\theta=0,\pi\), and vanishes at \(\theta=\pm\pi/2\), where the two replicas become equal in weight. Similarly, the \(n=\pm1\) Floquet replicas also acquire an asymmetry due to interference between different modulation pathways. Keeping the leading contributions,
\begin{align}
C_{+1}
&\simeq
J_1(\beta_1)J_0(\beta_2)
-
J_1(\beta_1)J_1(\beta_2)e^{-i\theta},
\\
C_{-1}
&\simeq
-
J_1(\beta_1)J_0(\beta_2)
-
J_1(\beta_1)J_1(\beta_2)e^{i\theta},
\end{align}
which leads to
\begin{equation}
|C_{+1}|^2-|C_{-1}|^2
\propto
\cos\theta. \label{s21}
\end{equation}
Thus, the \(+1\) and \(-1\) Floquet sidebands also exhibit a phase-dependent asymmetry. More generally, higher-order Floquet replicas can similarly acquire phase-dependent asymmetries due to interference among multiple Floquet pathways.

\subsection{Two-Tone Floquet Driving of the Cavity--Magnon Hybrid System}

In this section we extend the Floquet formalism to the driven cavity--magnon system in the bare $(m,c)$ basis, where the periodic modulation acts directly on the magnon frequency and the cavity couples to the magnon through the static interaction $g$.
 The Hamiltonian reads as
\begin{equation}
H
=
\omega_m m^\dagger m
+ \omega_c c^\dagger c
+ g\,(c^\dagger m + c m^\dagger)
+\Big[\delta\omega_1\cos(\Omega_1 t)+\delta\omega_2\cos(2\Omega_1 t+\theta)\Big] m^\dagger m , \label{eqs18}
\end{equation}
where $c$ is the cavity annihilation operator, and $\omega_c$ the bare cavity frequency.
To incorporate dissipation, we work with the non-Hermitian equations of motion for the bare mode amplitudes,
\begin{equation}
\frac{d}{dt}
\begin{bmatrix}
m\\
c
\end{bmatrix}
=
-i
\begin{bmatrix}
\omega_m-i\kappa_m + F(t) & g\\
g & \omega_c-i\kappa_c
\end{bmatrix}
\begin{bmatrix}
m\\
c
\end{bmatrix}
+
\sqrt{\kappa_{\rm ext}}
\begin{bmatrix}
m^{\rm in}\\
0
\end{bmatrix},
\label{eq:bare_eom}
\end{equation}
where
\begin{equation}
F(t)=\delta\omega_1\cos(\Omega_1 t)+\delta\omega_2\cos(2\Omega_1 t+\theta),
\end{equation}
with \(\kappa_c\) being the cavity damping rate. In the absence of the external modulation drives, the non-Hermitian cavity--magnon Hamiltonian matrix in Eq.~(\ref{eq:bare_eom}) can be diagonalized to obtain two hybrid eigenmodes $a$ and $b$ with complex eigenfrequencies $\tilde{\omega}_{a,b}=\frac{\omega_c+\omega_m - i(\kappa_c+\kappa_m)}{2}\pm \frac{1}{2}\sqrt{\left[(\omega_c-\omega_m)-i(\kappa_c-\kappa_m)\right]^2 + 4g^2}$, whose real (imaginary) parts give the resonance frequencies (linewidths) of the hybrid modes. A convenient description of the driven dynamics is obtained by expanding the bare-mode amplitudes in Floquet harmonics as
\begin{equation}
\begin{bmatrix}
m(t)\\
c(t)
\end{bmatrix}
=
\sum_{n\in\mathbb Z}
\begin{bmatrix}
m_n\\
c_n
\end{bmatrix}
e^{-in\Omega_1 t}.
\label{eq:bare_floquet_expansion}
\end{equation}
This expansion maps the problem onto a frequency lattice in which each Floquet sector \(n\) carries a two-component internal degree of freedom corresponding to the bare magnon and cavity amplitudes. The static magnon--cavity coupling \(g\) acts within each sector, while the periodic modulation couples different Floquet sectors only through the magnon component, generating nearest-neighbor hopping set by \(\delta\omega_1\) and next-nearest-neighbor hopping set by \(\delta\omega_2 e^{\pm i\theta}\). In the frequency domain, the steady-state Floquet amplitudes satisfy
\begin{equation}
\mathbf D(\omega)
=
-i\Big[\mathbf H_F-\omega\mathbf I\Big]^{-1}
\sqrt{\kappa_{\rm ext}} m^{in}\,\mathbf F,
\end{equation}
where \(\mathbf D(\omega)\) is the Floquet response vector and \(\tilde{\mathbf G}(\omega)=[\mathbf H_F-\omega\mathbf I]^{-1}\) is the Floquet Green function. In the truncated basis \(n=-N,\dots,N\), we order the solution as
\begin{equation}
\mathbf D(\omega)=
\begin{bmatrix}
\vdots\\
m_{-1}(\omega)\\
c_{-1}(\omega)\\
m_{0}(\omega)\\
c_{0}(\omega)\\
m_{1}(\omega)\\
c_{1}(\omega)\\
\vdots
\end{bmatrix},
\end{equation}
The drive vector $\mathbf F$ is a unit vector whose only nonzero component corresponds to the central magnon harmonic (i.e., the $n=0$ Floquet component), indicating that the external input couples exclusively to that mode. The corresponding Floquet Hamiltonian has block structure
\begin{equation}
(\mathbf H_F)_{n,n}=
\begin{bmatrix}
\omega_m-i\kappa_m-n\Omega_1 & g\\
g & \omega_c-i\kappa_c-n\Omega_1
\end{bmatrix},
\end{equation}
with nearest-neighbor Floquet couplings
\begin{equation}
(\mathbf H_F)_{n,n\pm1}
=
\frac{\delta\omega_1}{2}
\begin{bmatrix}
1 & 0\\
0 & 0
\end{bmatrix},
\end{equation}
and next-nearest-neighbor Floquet couplings
\begin{equation}
(\mathbf H_F)_{n,n\pm2}
=
\frac{\delta\omega_2}{2}e^{\pm i\theta}
\begin{bmatrix}
1 & 0\\
0 & 0
\end{bmatrix}.
\end{equation}
Thus, the modulation generates hopping between Floquet sectors only along the magnon leg of the synthetic lattice, while the cavity couples to the magnon locally within each sector through \(g\). The measured reflection probes the response at the incident frequency. In the Floquet description, the relevant quantity is therefore the zero-harmonic magnon amplitude \(m_0(\omega)\) and the reflection coefficient then follows as Eq.~(\ref{eq:reflection}).

\subsection{Derivation of the effective Floquet coupling}

To make the physical origin of the coupling asymmetry more transparent, we transform the Hamiltonian in Eq.~(\ref{eqs18}) into a rotating frame that removes the explicit time dependence from the magnon frequency. We introduce the unitary transformation
\begin{equation}
U(t)=e^{i\epsilon(t)m^\dagger m},
\end{equation}
under which the Hamiltonian transforms as
\begin{equation}
H_{\rm rot}(t)=UHU^\dagger-iU\partial_tU^\dagger .
\end{equation}
Using
\begin{equation}
U m U^\dagger = m e^{-i\epsilon(t)},
\qquad
U m^\dagger U^\dagger = m^\dagger e^{i\epsilon(t)},
\end{equation}
together with
\begin{equation}
-iU\partial_tU^\dagger
=
-\dot{\epsilon}(t)m^\dagger m,
\end{equation}
and noting that \(F(t)=\dot{\epsilon}(t)\), the explicit modulation term is exactly canceled by the inertial contribution generated by the transformation. The transformed Hamiltonian therefore becomes
\begin{equation}
H_{\rm rot}(t)
=
\omega_m m^\dagger m
+\omega_c c^\dagger c
+
g\left[
c^\dagger m e^{-i\epsilon(t)}
+
m^\dagger c e^{i\epsilon(t)}
\right].
\label{eq:Hrot_final}
\end{equation}
Thus, the periodic modulation no longer appears as a time-dependent diagonal frequency shift, but instead enters through a dynamical phase dressing the cavity--magnon coupling. Expanding the phase factor using Eq.~(\ref{s14}), the interaction Hamiltonian becomes
\begin{equation}
H_{\rm int}(t)
=
g\sum_{n\in\mathbb Z}
\left(
C_n c^\dagger m e^{-in\Omega_1 t}
+
C_n^* m^\dagger c e^{in\Omega_1 t}
\right).
\label{eq:Floquet_coupling}
\end{equation}
Substituting the Floquet expansions
\begin{equation}
m(t)=\sum_n m_n e^{-in\Omega_1 t},
\qquad
c(t)=\sum_n c_n e^{-in\Omega_1 t},
\end{equation}
into Eq.~(\ref{eq:Floquet_coupling}) shows explicitly that the cavity Floquet sector \(c_k\) couples to the magnon Floquet sector \(m_{k+n}\) with amplitude proportional to \(gC_n\). Thus, the replica-assisted transitions \(c_0\leftrightarrow m_{\pm1}\) are governed by \(gC_{\pm1}\), while the transitions \(c_0\leftrightarrow m_{\pm2}\) are governed by \(gC_{\pm2}\). Equivalently, the cavity replicas \(c_{\pm1}\) and \(c_{\pm2}\) couple back to the central magnon component \(m_0\) through the same coefficients \(C_{\mp1}\) and \(C_{\mp2}\), respectively. Therefore, the Floquet coefficients \(C_n\) directly determine the strength of the replica-assisted cavity--magnon hybridization processes, i.e., $\tilde g_n \sim gC_n$. Since both the \(n=\pm1\) and \(n=\pm2\) Floquet replicas exhibit a phase-dependent imbalance between \(C_{+n}\) and \(C_{-n}\), the corresponding replica-assisted coupling processes also become asymmetric. In particular, Eqs.~(\ref{s20}-\ref{s21}) show that, for the lowest-order sidebands considered here, the leading imbalance varies approximately as \(\cos\theta\). Consequently, the effective couplings associated with opposite Floquet sectors satisfy
\begin{equation}
\tilde g_{+n}\neq \tilde g_{-n},
\end{equation}
for \(n=\pm1,\pm2\), with the relative phase \(\theta\) directly controlling the asymmetry between opposite Floquet-assisted cavity--magnon coupling processes. More generally, higher-order Floquet sidebands are also expected to exhibit phase-dependent asymmetries due to interference among multiple Floquet pathways. However, unlike the \(n=\pm1\) and \(n=\pm2\) cases discussed above, the phase dependence of the higher-order coefficients can involve contributions from several competing interference processes and is not explored further here. The relative phase \(\theta\) therefore directly controls the asymmetry between opposite Floquet-assisted cavity--magnon coupling processes. At \(\theta=\pm\pi/2\), the interference term vanishes and the two couplings become equal, recovering a symmetric response similar to the single-tone case. In contrast, at \(\theta=0\) or \(\pi\), the interference is maximal, leading to strongly asymmetric effective couplings between opposite Floquet sectors. The asymmetry can also be understood in the hybridized cavity--magnon basis. 
In the absence of modulation, the hybrid modes are linear combinations of the bare magnon and cavity modes,
\begin{equation}
a=u m+v c,
\qquad
b=v m-u c ,
\end{equation}
with inverse relations \(m=u a+v b\) and \(c=v a-u b\), assuming \(u^2+v^2=1\). Substituting these relations into Eq.~(\ref{eq:Floquet_coupling}) generates both diagonal modulation terms and replica-assisted intermode coupling terms between the hybrid modes. In particular, the intermode part contains
\begin{equation}
\begin{aligned}
H_{\rm int}^{ab}(t)
=
g\sum_n
\Big[
&\left(v^2 C_n e^{-in\Omega_1 t}
-u^2 C_n^* e^{in\Omega_1 t}\right)a^\dagger b \\
&+
\left(-u^2 C_n e^{-in\Omega_1 t}
+v^2 C_n^* e^{in\Omega_1 t}\right)b^\dagger a
\Big].
\end{aligned}
\end{equation}
Thus, the hybrid-mode coupling inherits the phase-dependent imbalance of the bare Floquet coefficients, although the precise effective coupling is weighted by the magnon--cavity composition of the hybrid modes. 

\section{Experimental Details}

\subsection{Experimental Setup}

\begin{figure}[b]
\includegraphics[width=0.75\linewidth]{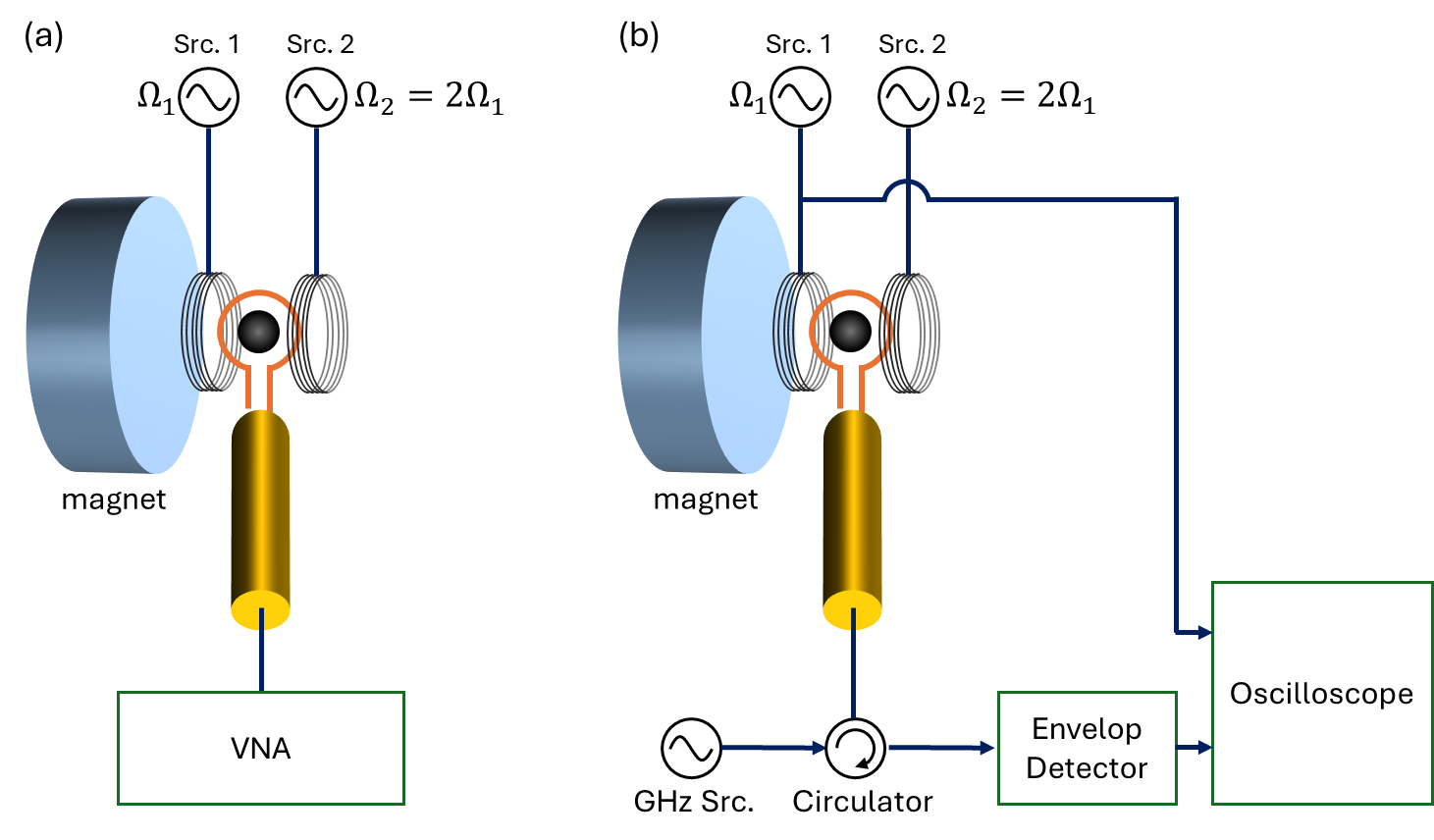}
\caption{Schematics of the experimental setup for (a) VNA spectrum measurement, and (b) time-resolved spectroscopy measurement, both for the magnon mode only. Src.: source.}
\label{figSM1}
\end{figure}

Our measurement setups are schematically shown in Fig.~\ref{figSM1}. A pair of small coils are used to provide the two Floquet drives. Both coils are compact and placed in close proximity to the YIG sphere to ensure high driving efficiency. The two signal sources are synchronized and phase locked to ensure a fixed phase relation. For the magnon reflection measurement, the coaxial probe is directly connected to a vector network analyzer (VNA) for $S_{11}$ measurement. For time-resolved spectroscopy, the magnon is excited using a GHz microwave source, and the reflected signal is measured using an oscilloscope after passing through an envelope detector.

\subsection{Time-Resolved Spectra}

\begin{figure}[hbt]
\includegraphics[width=0.99\linewidth]{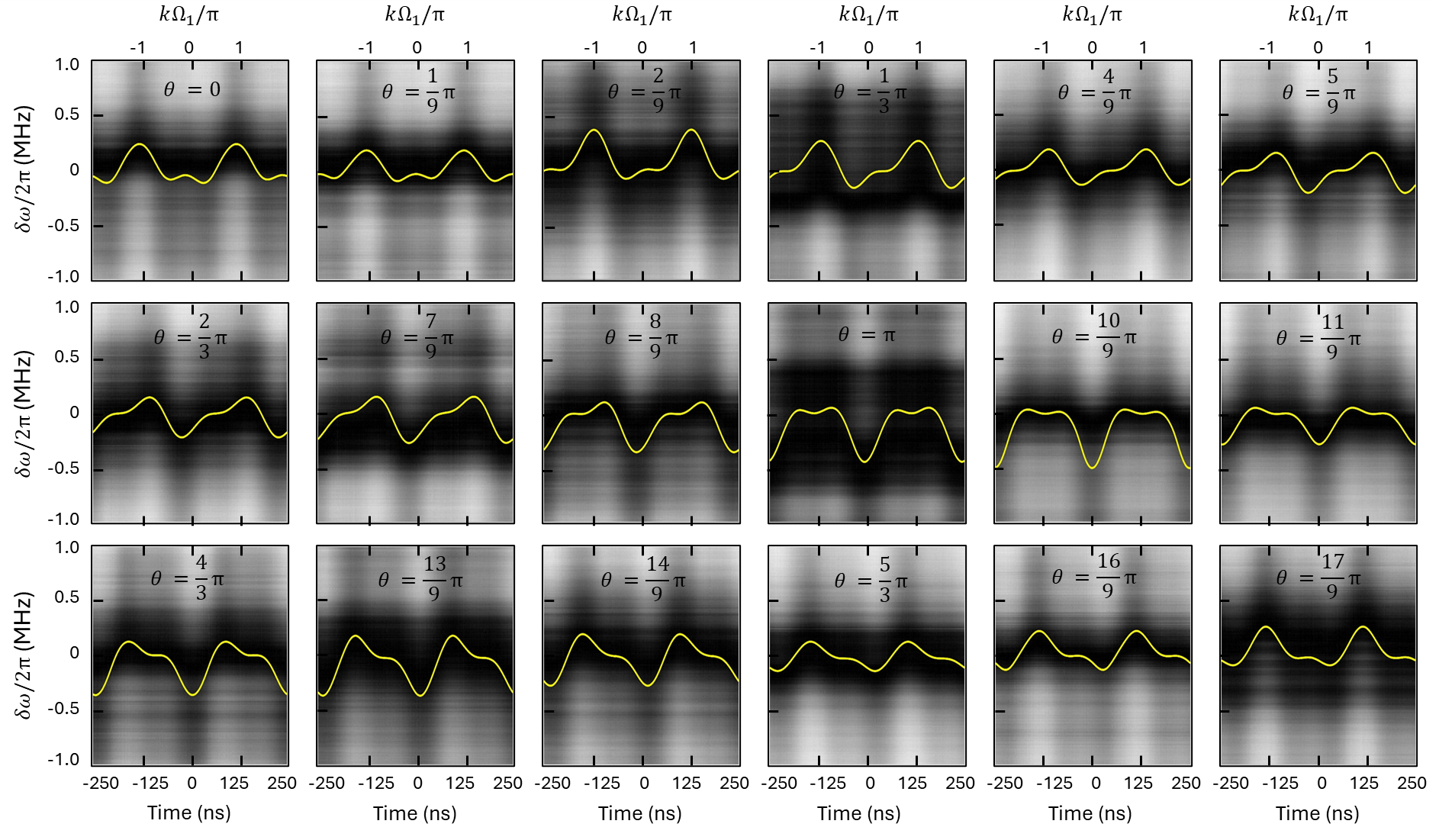}
\caption{Time-resolved spectra at different $\theta$ values. All other parameters are the same as Fig.2 of the main text.}
\label{figSM2}
\end{figure}

We performed time-resolved spectra for a wide range of $\theta$ values, and the results are shown in Fig.~\ref{figSM2}. To better visualize the variation of the magnon mode, whose linewidth is relatively large compared to the modulation depth, the center frequencies of the magnon modes are extracted and plotted as the yellow curves.

\subsection{Frequency Drift Correction}

\begin{figure}[hbt]
\includegraphics[width=0.7\linewidth]{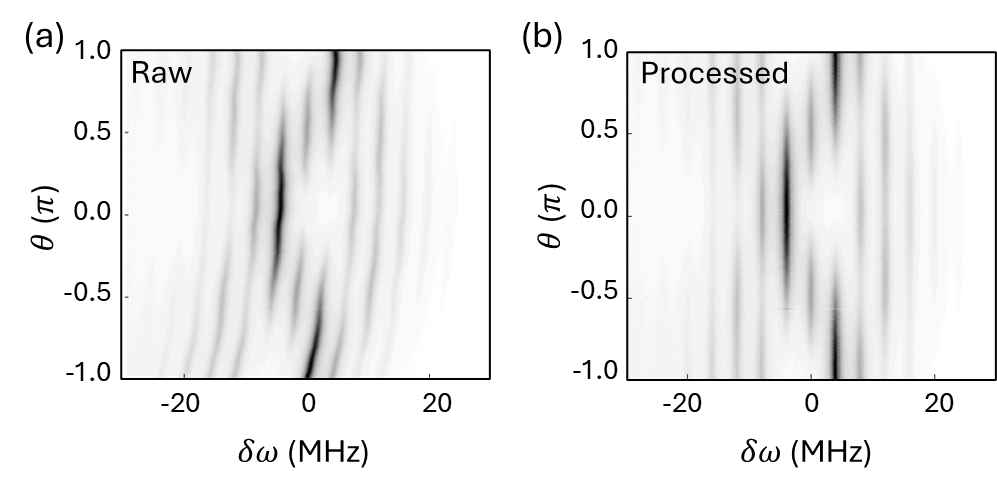}
\caption{(a) Magnon reflection spectrum measured using the VNA. (b) Processed magnon reflection spectrum by aligning all the magnon frequencies to remove the random frequency fluctuation induced by thermal effects or system stability issue, better revealing the variation of the Floquet replica as a function of $\theta$. Here $\delta\omega=\omega-\omega_0$ with $\omega_0/2\pi=4.0172$ GHz.}
\label{figSM3}
\end{figure}

In our experiment, the device configuration (such as coaxial probe positions) may vary slightly between measurements, causing minor variations in system parameters such as cavity frequency and linewidth that do not affect the core experimental observations. Additionally, when the system parameters are scanned during magnon reflection spectrum measurements, the magnon resonance frequency and corresponding Floquet replicas exhibit slight drift due to thermal effects and experimental stability limitations, as shown in Fig.~\ref{figSM3}(a). To suppress this frequency drift and clearly illustrate the Floquet replica dependence on system parameters, the measured spectra are processed by aligning the magnon mode frequency (i.e., the 0-th order Floquet replica) across all spectra. When the 0-th order replica is suppressed, another replica (e.g., 1-st order replica) is used for the alignment. This processing affects only the absolute frequency; the linewidth, extinction ratio, and spacing between neighboring replicas remain unchanged. The processed spectra are shown in Fig.~\ref{figSM3}(b), which corresponds to Fig.~3(a) in the main text.

\subsection{Asymmetry Factor Measurements}

\begin{figure}[hbt]
\includegraphics[width=1\linewidth]{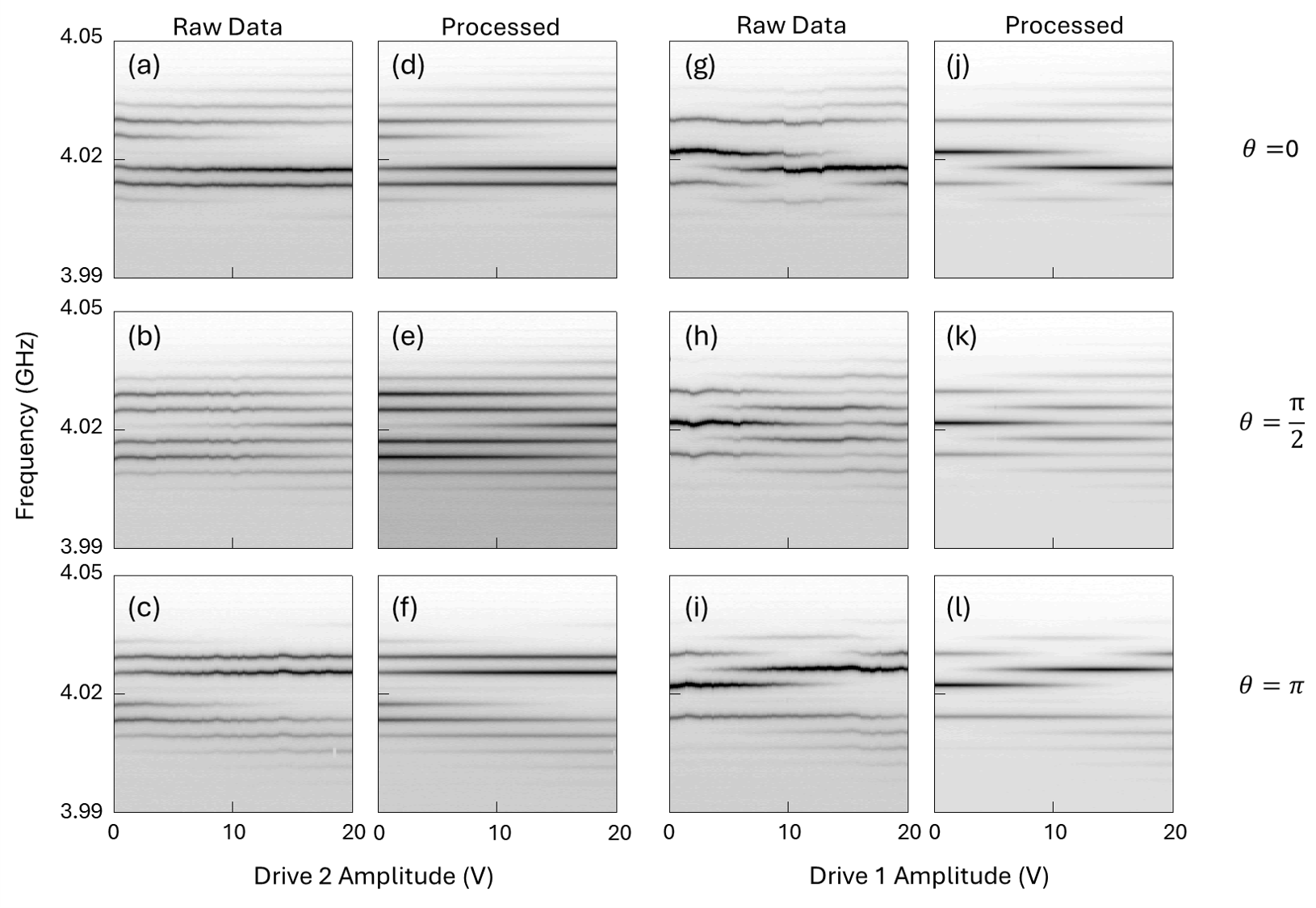}
\caption{Experimental reflection spectrum of the magnon mode for different phase angles: $\theta=0$ (top), $\pi/2$ (middle), and $\pi$ (bottom row). (a)-(c): Raw spectrum measured when $V_2$ is swept from 0 to 20 V while $V_1$ is fixed at $20~V$. (d)-(f) Processed spectrum by aligning the center frequencies of the magnon mode and the Floquet replicas in (a)-(c). (g)-(i): Raw spectrum measured when $V_1$ is swept from 0 to 20 V while $V_2$ is fixed at $20~V$. (j)-(l) Processed spectrum by aligning the center frequencies of the magnon mode and the Floquet replicas in (g)-(i).}
\label{figSM4}
\end{figure}

When measuring the asymmetry factor $\eta$, magnon reflection spectra are obtained at different driving amplitudes. As an example, Fig.~\ref{figSM4} plots the magnon reflection spectra when $V_2$ ($V_1$) is swept while $V_1$ ($V_2$) is fixed at 20 V, at three selected phase angles: $\theta=0$, $\pi/2$, and $\pi$. Evidently, the Floquet replicas are always symmetric for $\theta=\pi/2$, while the asymmetry level increases as $V_2$ ($V_1$) increases from 0 to 20 V for $\theta=0$ or $\pi$, which gives rise to the 2D map of the $\eta$ distribution in Fig.~3(c) of the main text.

\subsection{Continuous Scan of the Drive-Frequency Ratio}

\begin{figure}[hbt]
\includegraphics[width=0.99\linewidth]{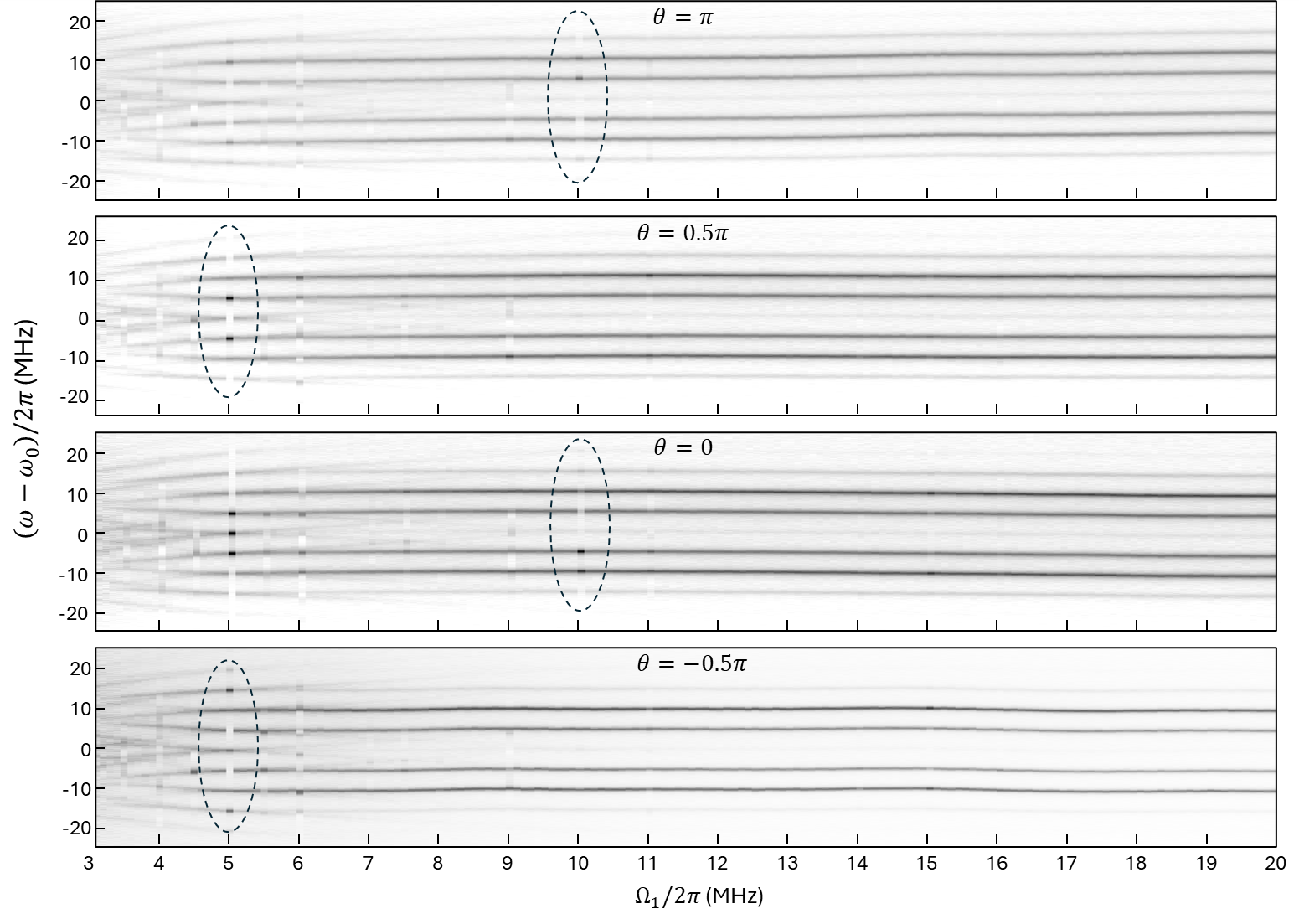}
\caption{Continuous scan of the dual-tone Floquet response versus drive frequency $\Omega_1$ at four relative phases $\theta = -0.5\pi, 0, 0.5\pi, \pi$. $\Omega_2/2\pi = 5$~MHz, both drives at 20~Vpp. Phase-dependent features appear at low-order commensurate ratios: 1:1 ($\Omega_1/2\pi = 5$~MHz), where the two tones interfere at the carrier, and 1:2 ($\Omega_1/2\pi = 10$~MHz), where asymmetric Floquet replicas appear at $\delta\omega/2\pi = \pm 10$~MHz with $\cos\theta$ phase dependence consistent with Fig.~3 of the main text. Key features are highlighted by the oval regions. Non-commensurate drive ratios produce no phase-controlled response.}
\label{figSM5}
\end{figure}

To investigate the generality of the dual-tone Floquet mechanism beyond the 1:2 drive ratio used in the main text, we performed continuous scans of the drive-frequency ratio $\Omega_1/\Omega_2$ with $\Omega_2/2\pi$ fixed at 5~MHz and $\Omega_1/2\pi$ swept from 3 to 20~MHz, at four representative relative phases $\theta = -0.5\pi, 0, 0.5\pi, \pi$. Both drives are set to 20~Vpp. The results are shown in Fig.~\ref{figSM5}.

Two distinct features are visible in the scan. Near $\Omega_1/2\pi = 5$~MHz (1:1 ratio), the two tones interfere directly at the carrier frequency: at $\theta = 0.5\pi$ the carrier is suppressed, while at $\theta = -0.5\pi$ it is enhanced, with intermediate behavior at $\theta = 0$ and $\pi$. Near $\Omega_1/2\pi = 10$~MHz (1:2 ratio), asymmetric Floquet replica generation appears at $\delta\omega/2\pi = \pm 10$~MHz, with the lower replica enhanced at $\theta = 0$ and the upper replica enhanced at $\theta = \pi$, consistent with the $\cos\theta$ phase dependence characterized in Fig.~3 of the main text. The asymmetry vanishes at $\theta = \pm 0.5\pi$, as expected. The 1:2 feature persists at neighboring low-order commensurabilities such as 5:9 and 5:11, and weaker phase-dependent features are also visible at higher-order commensurabilities such as 1:3 ($\Omega_1/2\pi = 15$~MHz). Ratios outside these low-order-commensurate points---for example, very-high-order commensurabilities such as 10:19 or 25:49---produce no observable phase-controlled features, reflecting the requirement that the dual-tone interference involve a low-order multiphoton process for measurable contrast.

These results confirm that the phase-controlled asymmetric replica generation is a specific feature of the 1:2 commensurate condition, and that the dual-tone Floquet method produces distinct phenomenology at different commensurate ratios.

\subsection{Autler-Townes Splitting in the Hybrid System}

\begin{figure}[b]
\includegraphics[width=0.99\linewidth]{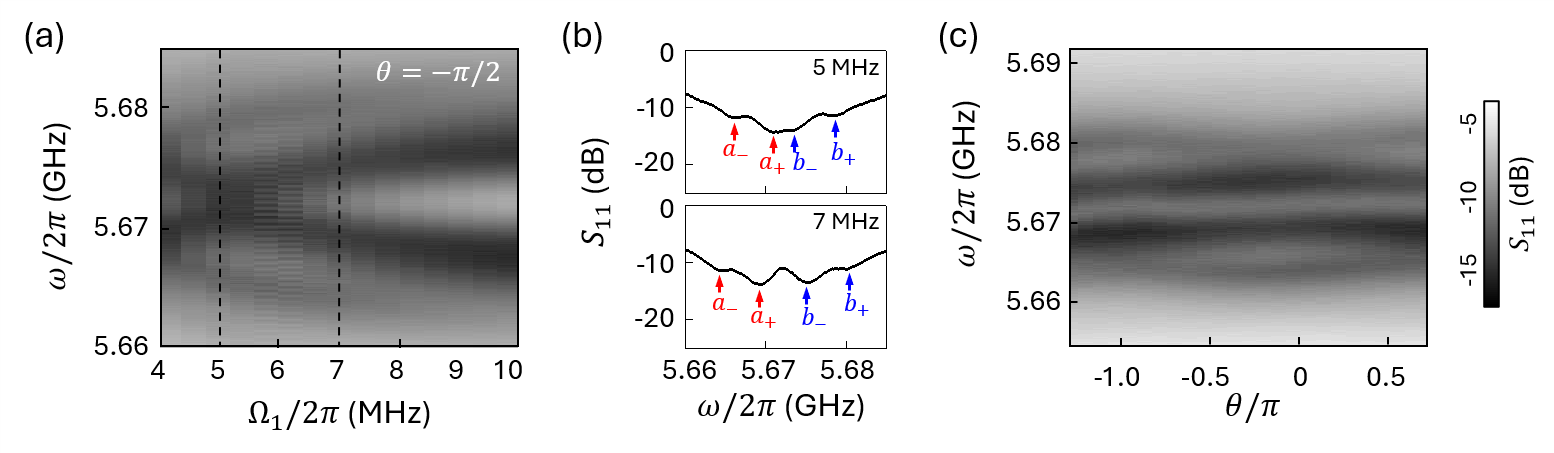}
\caption{Additional hybrid system measurements complementing Fig.~4 of the main text. (a) Reflection spectrum as a function of drive frequency $\Omega_1$ at $\theta = -\pi/2$, showing symmetric ATS on both hybrid modes. Dashed lines indicate $\Omega_1/2\pi = 5$~MHz and 7~MHz. (b) Linecuts at $\Omega_1/2\pi = 5$~MHz (top) and 7~MHz (bottom) at $\theta = -\pi/2$, with hybrid mode branches labeled ($a_-$, $a_+$, $b_-$, $b_+$). (c) Reflection spectra as a function of drive phase $\theta$ at $\Omega_1/2\pi = 7$~MHz, showing the continuous evolution of the asymmetric hybridization between the two hybrid modes with reduced spectral overlap compared to the 5~MHz case shown in Fig.~4(b) of the main text.}
\label{figSM6}
\end{figure}

In Fig.~4 of the main text, the spectrum is dominated by the ATS arising from the $n=\pm 2$ sidebands, while the off-resonant $n=\pm 1$ sidebands are significantly weaker and become indistinguishable. These, together with higher-order sidebands (e.g., $n=\pm 3$), contribute faint features to the background, complicating the measured spectrum. In addition, experimental background effects, such as cable interference, can introduce visual artifacts, further obscuring the identification of the phase-dependent evolution of the ATS. Nevertheless, this phase-dependent ATS variation is unambiguously revealed in Fig.~4(b) when the phase is continuously scanned.

It is worth noting that since the ATS of both modes $a$ and $b$ are comparable to their spectral separation, the inner branches $a_+$ and $b_-$ merge into a single unresolved dip. In the extraction of $\eta_\mathrm{ATS}$, this merged dip is treated as one feature at its center frequency, which approximates the average of $\omega_{a_+}$ and $\omega_{b_-}$ (frequencies of $a_+$ and $b_-$, respectively), causing both ATS values to be slightly larger than their actual values. As a result, the extracted $\eta_\mathrm{ATS}$ systematically underestimates the actual coupling asymmetry, while its qualitative $\cos\theta$ dependence remains unaffected.

Figure~\ref{figSM6} shows additional reflection spectra from the hybrid cavity--magnon device complementing Fig.~4 of the main text. Figure~\ref{figSM6}(a) is identical to the middle panel of Fig.~4(a) in the main text, showing the reflection spectrum as a function of drive frequency $\Omega_1$ at $\theta = -\pi/2$, which exhibits symmetric ATS on both modes. Figure~\ref{figSM6}(b) shows linecuts of the reflection spectra at $\Omega_1/2\pi = 5$~MHz (top) and 7~MHz (bottom), as indicated by the two vertical dashed lines in Fig.~\ref{figSM6}(a). The hybrid modes ($a_-$, $a_+$, $b_-$, $b_+$) originated from the Floquet-mediated mode coupling are labeled. The mode splittings are clearly resolved at both drive frequencies. In the 5 MHz case, the two inner hybrid modes $a_+$ and $b_-$ are very close in frequency, with a spacing smaller than their linewidths, and thus are barely distinguishable.

Figure~\ref{figSM6}(c) shows the phase-sweep measurement at $\Omega_1/2\pi = 7$~MHz, complementing the 5~MHz phase sweep shown in Fig.~4(d) of the main text. At this off-resonant drive frequency ($2\Omega_1/2\pi = 14$~MHz $\neq \Delta/2\pi = 10$~MHz), the induced splitting is smaller than at 5~MHz due to the reduced drive amplitude from the coil's low-pass response, resulting in reduced spectral overlap between the ATS features of the two modes. In this case, the two inner modes, $a_+$ and $b_-$, are well separated and clearly distinguishable. This allows the periodic switching of ATS between single-sided and double-sided as $\theta$ is varied to be more clearly resolved, confirming the continuous phase-controlled asymmetry of the two-mode coupling demonstrated in the main text.

\end{document}